\documentclass[twocolumn,preprintnumbers,secnumarabic,amsmath,amssymb,superscriptaddress,nofootinbib]{revtex4}

\usepackage{graphicx}
\usepackage{xcolor}
\usepackage{dcolumn}
\usepackage{bm}
\definecolor{linkcolor}{rgb}{0.0,0.3,0.5}
\usepackage[unicode, colorlinks=true, linkcolor=linkcolor, citecolor=linkcolor, 
filecolor=linkcolor,urlcolor=linkcolor, pdfusetitle]{hyperref}

\usepackage{makecell}
\usepackage{nccmath}

\usepackage{stackengine}

\newcommand{\nn}{\nonumber}

\newcommand{\beq}{\begin{equation}}
\newcommand{\eeq}{\end{equation}}

\newcommand{\bea}{\begin{eqnarray}}
\newcommand{\eea}{\end{eqnarray}}

\def\be{\begin{equation}}
\def\ee{\end{equation}}
\def\bal{\begin{align}}
\def\eal{\end{align}}
\def\nn{\nonumber}

\def\({\left(}
\def\){\right)}
\def\[{\left[}
\def\]{\right]}

\usepackage{epsf}
\usepackage{fancyhdr}
\usepackage{bm}
\usepackage{enumerate} 

\usepackage{empheq} \empheqset{box=\fbox}

\usepackage{graphicx}
\usepackage{tikz}
\usetikzlibrary{decorations.pathreplacing,calligraphy} 
\usetikzlibrary{decorations.markings}
\tikzset{->-/.style={decoration={
			markings,
			mark=at position #1 with {\arrow{stealth}}},postaction={decorate}}}
\usepackage{adjustbox}
\usepackage{pgfplots}
\pgfplotsset{compat=1.11}
\usepgfplotslibrary{fillbetween}
\usetikzlibrary{intersections}
\usetikzlibrary{patterns}

\pgfdeclarelayer{bg}
\pgfsetlayers{bg,main}

\begin{document}

\setlength{\abovedisplayskip}{4pt}
\setlength{\belowdisplayskip}{4pt}
\setlength{\intextsep}{3pt}
\setlength{\footnotesep}{0pt}
\setlength{\skip\footins}{1cm}

\title{ KKLT Ex Nihilo}

\vspace*{-.1cm}

\author{Iosif Bena}
\affiliation{{Institut de Physique Th\'eorique,
	Universit\'e Paris Saclay, CEA, CNRS,
 F-91191 Gif-sur-Yvette, France
}}
\author{Yixuan Li}
\affiliation{{Dipartimento di Fisica e Astronomia “Galileo Galilei”, Università di Padova, and}}
\affiliation{{INFN, Sezione di Padova, Via Marzolo 8, 35131 Padova, Italy}}

\author{Severin L\"ust}
\affiliation{{Laboratoire Charles Coulomb (L2C), Universit\'e de Montpellier, CNRS, F-34095, Montpellier, France}}

\begin{abstract}
\noindent
Flux compactifications that give three- or four-dimensional Anti de Sitter vacua with a parametrically small negative cosmological constant are claimed to be ubiquitous in String Theory. However, the 1+1 and 2+1 dimensional CFT duals to such vacua should have very large central charges and rather unusual properties.
We construct brane configurations that source these would-be AdS flux compactifications, and identify certain UV AdS geometries that these branes source. The central charge of the CFT duals to these UV AdS geometries place lower bounds on the absolute values of the cosmological constants of the AdS vacua.
These bounds are incompatible with the scale-separation needed to construct realistic cosmological models. 
\end{abstract}

\maketitle

\section{Introduction}
\vspace*{-.45cm}

The expansion of our universe is accelerating, and the best explanation for this acceleration is a small positive cosmological constant. However, String Theory, which is the leading candidate for a theory that unifies all interactions and which has no trouble producing all ingredients of the Standard Model, appears to have a hard time producing solutions with a small and positive cosmological constant. There are theorems that de Sitter vacua cannot be obtained from non-singular supergravity compactifications without relying on intrinsically string-theoretical effects
\cite{Maldacena:2000mw}, and there are also bottom-up swampland conjectures that such solutions are not stable \cite{Obied:2018sgi,Ooguri:2018wrx}.

The most popular and well-controlled scenario for constructing de Sitter vacua in String Theory \cite{Kachru:2003aw} consists of first constructing Anti-de Sitter solutions with a cosmological constant that is exponentially suppressed in the ratio between the compactification scale and the Planck scale,
and then adding anti-branes to uplift this small negative cosmological constant to a positive one. The Anti-de Sitter solutions are obtained by adding fluxes and branes to give mass to the fields corresponding to the flat directions (moduli) of the compactification manifold and, according to the KKLT proposal and the several hundred other papers that investigated this issue, they can be scale separated: their cosmological constant can be parametrically smaller than the size of the compactification manifold.

By the AdS-CFT correspondence \cite{Maldacena:1997re}, such AdS vacua are dual to conformal field theories that have an exponentially large central charge and an unusual spectrum of operators \cite{Conlon:2021cjk,Apers:2022tfm}. There are several hints based on CFT bootstrap techniques \cite{Baume:2020dqd, Perlmutter:2020buo, Bobev:2023dwx, Perlmutter:2024noo} that such CFT's may be problematic, but no definitive argument ruling them out. Furthermore, there are also swampland-type conjectures that scale-separated AdS compactifications cannot be constructed \cite{Lust:2019zwm}, or at least must be very constrained \cite{Li:2023gtt,Palti:2024voy}.

An interesting new approach to settle the question of how much scale separation can exist in flux compactifications has been pioneered by Vafa, Wiesner, Xu and one of the authors \cite{Lust:2022lfc}. They argued that one can construct a 2+1-dimensional {\em KKLT domain wall}, that contains D5 and NS5 branes that have 2+1 common directions and wrap the 3-cycles dual to the fluxes of the KKLT solution.
They also argued that the number of degrees of freedom on the domain wall, $c_{\mathrm{UV}}$, of order the product of the number of D5 and NS5 branes, is larger than the central charge of the scale-separated AdS solution, $c_{\mathrm{IR}}$:
\be \label{LVWX_estimate_c_UV}
c_{\mathrm{IR}} \leq c_{\mathrm{UV}} \sim (N_5)^2 \,,
\ee
with $N_5$ the number of five-branes.
This is supposed to put an upper bound on the amount of possible scale separation, as $c_{\mathrm{IR}}$ measures the AdS radius in the lower-dimensional Planck units. 

However, there are several objections that one can bring to the arguments in \cite{Lust:2022lfc}. The first is that the central charge of a domain wall does not constrain the central charges on the theories on its two sides. The second is that the relation between the number of degrees of freedom of a system of branes and the central charge of a given AdS space can only be established when the AdS solution is sourced by this system of branes. The third is that the degrees of freedom of the branes of the domain wall are counted in \cite{Lust:2022lfc} at weak coupling, but as one increases the coupling, new non-perturbative degrees of freedom might appear which may drastically increase the central charge. 

In this letter we solve these three problems by constructing a brane system that can create the KKLT solution out of nothing (ex nihilo), and that can source an ``UV" AdS solution that flows in the infrared to the KKLT AdS solution. Such a brane system can either be constructed for KKLT AdS$_4$ vacua, obtained by compactifying Type IIB String Theory on six-dimensional Calabi-Yau (CY) manifolds with fluxes, or for putative scale-separated AdS$_3$ vacua obtained by compactifying M-theory on eight-dimensional CY manifolds with fluxes. The central charge of the UV near-brane AdS solution is guaranteed to capture all the degrees of freedom of the branes and, as we will see, is parametrically larger than the central charge calculated at weak coupling in \cite{Lust:2022lfc}.

We begin with the construction of the M-theory branes sourcing a putative scale-separated KKLT-like AdS$_3$ solution, for which the UV AdS$_3$ solution is easy to construct. We then construct the UV AdS$_4$ solution near the D3, D5 and NS5 branes that source a putative  scale-separated AdS$_4$ compactifications, and calculate the maximal scale separation they allow.

\vspace*{-.75cm}
\section{M-theory compactifications to AdS\texorpdfstring{$_3$}{3}.}
\vspace*{-.5cm}

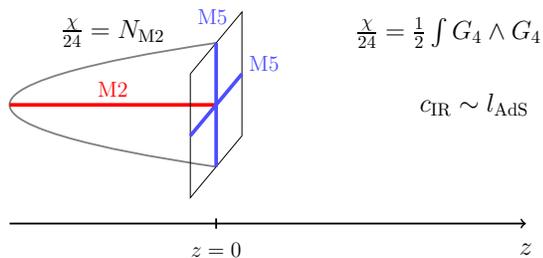
\begin{figure}[t]
	\begin{adjustbox}{max totalsize={0.4\textwidth}{\textheight},center}
		\begin{tikzpicture}
			\begin{scope}[shift={(0,0)}]
                    %
                   \draw[gray, line width = 1,rotate=-90] (-1.2,2) parabola bend (0,-2) (1.2,2);
                
				%
				%
				%
                \draw [red, line width=2] (-2,0) --(2,0);
				\node[red, align = center, centered] at (0, 0.3) { \large M2};
                \node[align = center, centered] at (0, 1.4) { \Large $\frac{\chi}{24}=N_\mathrm{M2}$};
				%
				%
                    %
				\draw [ black ] (1.5,0.6) --(2.5,1.8) -- (2.5,-0.6) -- (1.5,-1.8) -- cycle;
                    \draw [blue!70, line width=2] (1.5,-0.6) --(2.5,0.6);
                    \draw [blue!70, line width=2] (2,1.2) --(2,-1.2);
                    \node[blue!70, align = center, centered] at (3, 0.8) { \large M5 };
                    \node[blue!70, align = center, centered] at (2, 1.7) { \large M5 };
				%
                    %
				%
                    %
                    \node[align = center, centered] at (7, 0) { \Large $c_{\mathrm{IR}}\sim l_{\mathrm{AdS}}$};
                    \node[align = center, centered] at (6.5, 1.4) { \Large $\frac{\chi}{24}=\frac12 \int G_4\wedge G_4$};
                    %
                    %

				    \draw [ black, line width=1, ->] (-2, -2.3) --(8,-2.3);
                    \node[align = center, centered] at (8, -2.8) {\Large $z$};
                    \draw [ black, line width=1] (2, -2.2) --(2,-2.4);
                    \node[align = center, centered] at (2, -2.8) {\large $z=0$};

			\end{scope}

		\end{tikzpicture}
	\end{adjustbox}
	\vspace*{-.75cm}
	\caption{
		The domain wall with a contracting Calabi-Yau.
		}
	\label{fig:ex-nihilo_DW}
	\vspace*{-.5cm}
\end{figure}

M-theory compactified on an eight-dimensional Calabi-Yau manifold with self-dual four-form fluxes, $G_4$, is expected to yield an AdS$_3$ vacuum once non-perturbative effects are taken into account.
As explained in \cite{Lust:2022lfc,Minasian:1999gg,Silverstein:2003jp,deAlwis:2014wia}, the AdS$_3$ vacuum is sourced by a 1+1 dimensional domain wall that consists of M5 branes wrapping a special Lagrangian submanifold (SLag), $L_4$, Poincar\'e dual to $G_4$ in the CY. These M5 branes create jumps in the values of the $G_4$ fluxes.  However, the fluxes on both sides of the domain wall have to satisfy the tadpole cancellation condition, 
\be \label{tadpole_cancellation}
\frac{\chi_{\mathrm{CY}_4}}{24} = N_{\mathrm{M2}} + \frac12 \int_{\mathrm{CY}_4} G_4\wedge G_4 \,.
\ee
As shown in Fig.~\ref{fig:ex-nihilo_DW}, we consider a domain wall where the number of branes is such that all the fluxes of the AdS compactification on the right ($z>0$) jump to zero on the left ($z<0$). The flux-less CY$_4$
has a nontrivial Euler characteristic, $\chi_{\mathrm{CY}_4}$, and to satisfy \eqref{tadpole_cancellation}, there must be  $\chi_{\mathrm{CY}_4}/24$ M2 branes on the left of the domain wall. These M2 branes are points in the CY$_4$ and fill up the transverse (2+1)-dimensional space. They end on the M5 branes of the domain wall and pull them into spikes \cite{Bena:2022wpl,Eckardt:2023nmn}, similar to the Callan-Maldacena spike \cite{Callan:1997kz}.

This construction has two new characteristics.
First, because the CY$_4$ on the left has no fluxes, its moduli will receive quantum corrections and run. These corrections generically depend on all moduli, including the position of the M2 branes inside the CY$_4$. Generically we expect a minimum of the moduli potential somewhere in the deep interior of the moduli space (as also suggested long ago by Dine and Seiberg \cite{Dine:1985he}).
Therefore, we will assume without loss of generality that the CY$_4$ on the left will shrink to string size.
Hence, the spacetime will end on the left of our branes, and our domain wall with M2 brane spikes can be thought of as sourcing the AdS$_3$ solution.

Second, the M2 branes that form the spikes and the M5 branes in the domain wall give rise to a new AdS$_3$ geometry, which can be thought as the UV AdS that flows to the KKLT-like scale-separated AdS. There are many ways to arrange the M5 branes wrapping the SLag and the M2 branes ending on them. Since our goal is to place an upper bound on the central charge of the IR CFT, we will consider the arrangement that has the largest possible central charge: all the $(N_5)^2$ self-intersections of the M5 branes coincide and all the M2 branes end on this intersection.  

We can now zoom in on the self-intersection point (see Table \ref{tab:M5-M2_config}): The $N_5$ self-intersecting M5 branes appear locally like $N_5$ M5 branes wrapping the 1234 directions and $N_5$ M5 branes wrapping the 5678 directions as well as the direction of the domain wall, $y$. The M2 branes also wrap this direction, as well as the direction orthogonal to the domain wall, $z$.

\begin{table}[h]
    \centering
\begin{tabular}{|c|c|c|c|c|c|c|c|c|c|c|c|}
\hline
   & $t$   & $y$   & $z$                & 1   & 2   & 3   & 4   & 5 & 6 & 7 & 8 \\ \hline
M5 & $\otimes$ & $\otimes$ & $\stackrel{z=0}{\bullet}$ & $\otimes$ & $\otimes$ & $\otimes$ & $\otimes$ &   &   &   &    \\ \hline 
 M5' & $\otimes$ & $\otimes$ & $\stackrel{z=0}{\bullet}$ & & & & & $\otimes$ & $\otimes$ & $\otimes$ &$\otimes$ \\ \hline 
M2 & $\otimes$ & $\otimes$ & $\stackrel{z<0}{\otimes}$       &     &     &     &     &   &   &   &    \\ \hline
\end{tabular}
    \caption{The zoom-in on the branes of the brane system sourcing the putative scale-separated AdS$_3$ vaccuum.     }
    \label{tab:M5-M2_config}
\end{table}

The brane system of Table \ref{tab:M5-M2_config} generates two AdS regions. The first region is an $\mathrm{AdS}_3 \times \mathrm{CY}_4$ and is sourced by the M5-M5 domain wall far away from the branes. This KKLT-like AdS$_3$ is spanned by the coordinates $(t,y,z)$. 
The second AdS$_3$ region can be found in the near-brane limit of the M5-M5-M2 UV domain wall, and is spanned by $t, y$ and another coordinate that is a complicated function of $z$ and the radial directions inside the $1234$ and $5678$ planes, defined in equation \eqref{metricM5M5M2_general_ansatz}.

It is important to understand which of these AdS$_3$ spaces is the UV AdS$_3$ and which is the IR AdS$_3$. Because the branes sourcing the KKLT   AdS$_3$ live at the end of this space, one may naively think of them as end-of-the-world branes, like D3 branes on a Coulomb branch sourcing $\rm AdS_5 \times S^5$. However, this is not correct. There is a key difference between our branes and end-of-the-world branes: when one increases the charge of the end-of-the-world branes, the radius of the AdS space sourced by them increases. On the other hand, when one increases the charge of our branes, the radius of the KKLT-like AdS space decreases. 

One could also argue that if our branes were in the infrared of the KKLT-like AdS space this would violate the holographic principle. One can consider a solution with a domain wall of half the charges of our wall at $z=z_0>0$ in Fig.~\ref{fig:ex-nihilo_DW}, an intermediate AdS region between $z=0$ and $z=z_0$ and another end-of-spacetime wall at $z=0$ with the other half of the charges. The intermediate AdS, between $0<z<z_0$ would have a larger radius than the KKLT AdS, and hence must necessarily be in its UV. This demonstrates that the branes sourcing the KKLT AdS spacetime are in the UV. Hence, the near-brane AdS$_3$ they source is in the UV of the KKLT AdS$_3$, and its central charge places an upper bound on the central charge of the CFT dual to the KKLT AdS.

The supergravity solution sourced by the M2 and M5 branes in Table \ref{tab:M5-M2_config} is governed in general by a complicated Monge-Amp\`ere equation \cite{Lunin:2007mj,Bena:2023rzm}. However, in the infrared of this system one can find a warped solution of the form $\mathrm{AdS}_3 \times \mathrm{S}^3 \times \mathrm{S}^3 \times \Sigma_2$, where $\Sigma_2$ is a Riemann surface. These solutions have been constructed in \cite{Bachas:2013vza} and the relation between the Riemann-surface coordinates and the  coordinates in Table \ref{tab:M5-M2_config} has been worked out in \cite{Bena:2023rzm, Bena:2024dre}. To evaluate the central charge of the CFT dual to this warped AdS$_3$ one needs to specify the precise Riemann-surface singularities corresponding to M2 and M5 branes, and to compute the AdS radius in terms of the three-dimensional Planck length, which depends on the volume of  $\mathrm{S}^3 \times \mathrm{S}^3 \times \Sigma_2$. 

However, there is a simpler and more intuitive way to compute this central charge, by assuming the $z$ direction to be compact, and smearing the solution along $z$. As we will see, the central charge does not depend on the periodicity of $z$, which confirms the validity of this procedure.

\vspace*{-.75cm}
\subsection{A smeared M2-M5-M5' intersection}
\vspace*{-.45cm}

We will smear the M2-M5-M5' spikes in Table 1 along the M2 direction, $z$. The M5 branes are localized at the origin of the 3456 space ($r'=0$) and the M5' branes are localized at the origin of the 1234 space ($r=0$). Smearing these spikes along $z$ produces a $z$ independent solution, which can be shown \cite{Bena:2023rzm,Bena:2024dre} to be the one constructed in \cite{deBoer:1999gea}:
\begin{align} \label{metricM5M5M2_general_ansatz}
ds^2=&\bigl(H_T^{-1}Z_5Z_5'\bigr)^{2/3}\[ \bigl(Z_5Z_5'\bigr)^{-1}\(-dt^2+dy^2\)
+ dz^2 \] \nn \\
&+ H_T^{1/3}\bigl(Z_5\bigr)^{-1/3}\bigl(Z_5'\bigr)^{2/3} \bigl(dr^2+r^2d\Omega_{(1)}^2\bigr) \nn \\
&+ H_T^{1/3}\bigl(Z_5\bigr)^{2/3}\bigl(Z_5'\bigr)^{-1/3} \bigl(dr'^2+r'^2d\Omega_{(2)}^2\bigr) \,.
\end{align}

In this solution $Z_5$ and $Z_5'$ can be thought of as the harmonic functions associated to the two species of M5 branes,
\be
Z_5=1+\frac{l_{\mathrm{P}}^2 \, n_F^{(1)}}{r'^2} \,, \qquad Z_5'=1+\frac{l_{\mathrm{P}}^2 \, n_F^{(2)}}{r^2} \,,
\ee
and the function $H_T$ is associated to the M2-branes, but is not harmonic anymore
\be \label{HT_dBPS}
H_T= \(1+\frac{l_{\mathrm{P}}^2 \, n_T^{(1)}}{r'^2}\) \(1+\frac{l_{\mathrm{P}}^2 \, n_T^{(2)}}{r^2}\) \,.
\ee

As explained in \cite{deBoer:1999gea}, one can take a near-brane decoupling limit of this solution.%
\footnote{This limit is $l_{\mathrm{P}} \rightarrow 0$ with $U\equiv\frac{r^{2}}{l_{\mathrm{P}}^{3}}$ and $U'\equiv\frac{r'^{2}}{l_{\mathrm{P}}^{3}}$ held fixed. It is equivalent to dropping the 1's in the harmonic functions.} The metric becomes 
\begin{align} \label{metric_dBPS_NH_u_lambda}
{ds^2 \over l_\mathrm{P}^2}=&\left({u \over l_{\mathrm{AdS}}}\right)^2 (-dt^2 + dy^2) 
+ \(\frac{l_{\mathrm{AdS}}}{u^2}\)^2 du^2  \\ \nn 
& + \left(n_T n_F\right)^{\frac13} 
 \left( d\Omega^2 + d\Omega'^2 \right) 
 + d \lambda^2 +\left({n^2_F \over  n_T}\right)^{\frac23} dz^2 \,,
\end{align}
where
\begin{equation}
l_{\mathrm{AdS}} \equiv \tfrac{1}{\sqrt{2}} \left(n_T n_F\right)^{\frac16}
\end{equation}
is the dimensionless $\mathrm{AdS}_3$ radius. The coordinates $u$ and $\lambda$ are hyperbolic coordinates in the plane spanned by the two radial directions inside the M5 and M5' branes:
\be
u =  \frac{2 \sqrt{U U'}}{l_{\mathrm{AdS}}}\,, \qquad 
\lambda= \frac{l_{\mathrm{AdS}}}{2}  \log \frac{U}{U'} \,.
\ee

Even if $\lambda$ is a non-compact coordinate, we will formally compactify it on an interval of length $L_{\lambda}$ in order to compute the quantized charges and the central charge of the dual CFT. As we will see, the final result does not depend on $L_{\lambda}$, so we can safely send it to infinity and decompactify the $\lambda$ direction again. 

The dimensionless quantities $n_F^{(i)}$ and $n_T\equiv n_T^{(1)} n_T^{(2)}$ 
measure the M5 and M2 densities in 11d Planck units%
\footnote{Since our domain wall sources a KKLT-like solution that has self-dual fluxes ($G_4=\star G_4$), the number of M5 and M5' branes (and their densities) are equal: $n_F^{(1)} = n_F^{(2)} \equiv n_F$}
and can be expressed in terms of number of M5 and M2 branes:
\begin{align}
N_5 &= \frac{1}{(2\pi l_\mathrm{P})^3} \int_{ S_z \times  S^3_{(2)}} G_4
= \frac{L_z}{2\pi} n_F  \,, \\
N_2 &= \frac{1}{(2\pi l_\mathrm{P})^6} \int_{S_\lambda \times S^3_{(1)} \times S^3_{(2)}} G_7 \, 
= \frac{\sqrt{2}}{8\pi^2} L_\lambda \(n_T n_F\)^{5/6}    \,, \nn
\end{align}
where $G_7=\star G_4 - \frac{1}{2} C_3 \wedge G_4$. $L_z$ and $L_\lambda$ are the dimensionless periodicities of the coordinates $z$ and $\lambda$.

The central charge of the near-brane UV CFT can be calculated from the dimensionless AdS radius and the 3d Planck length and expressed in terms of the number of M2 and M5 branes
\begin{align}\label{eq:Mtheoryc}
c &=\frac{3}{2} \frac{l_{\rm P} l_{\mathrm{AdS}_3}}{G_\mathrm{N}^{(3)}}
= 3 N_2 N_5 \,.
\end{align}

Using the M2 tadpole condition  \eqref{tadpole_cancellation}, we can see that this central charge scales like $(N_5)^3$. This is a larger central charge than the one derived at weak coupling in \cite{Lust:2022lfc}, by considering brane deformations. This is not surprising: the brane system is more complicated, and the $AdS$ geometry takes into account both the perturbative degrees of freedom counted in \cite{Lust:2022lfc} and also the non-perturbative degrees of freedom.

\vspace*{-.75cm}
\section{IIB compactifications to ${\text{AdS}}_4$} \label{sec:IIB}
\vspace*{-.45cm}

As shown in Table \ref{tab:D5-NS5-D3_config}, the branes sourcing the analogous type IIB  ${\text{AdS}}_4$ vacuum are D5 and NS5 branes wrapping three-cycles on a CY three-fold, on which a number of spacetime-filling D3 branes end. The D5 and NS5 branes intersect at points in the CY manifold, and the way to obtain the largest number of degrees of freedom is to assume that they are all coincident and intersect at a single point, on which the $N_3$ D3 branes terminate as well. In the neighborhood of such an intersection, we can choose coordinates such that the two brane stacks are spanned by the coordinates $456$ and $789$, respectively.

\begin{table}[h]
    \centering
\begin{tabular}{|c|c|c|c|c|c|c|c|c|c|c|}
\hline
   & $t$   & $y_1$    &$y_2$& $z$                & 4   & 5   & 6   & 7 & 8 & 9 \\ \hline
D5& $\otimes$ & $\otimes$  &$\otimes$  & $\stackrel{z=0}{\bullet}$ & $\otimes$ & $\otimes$ & $\otimes$ &   &   &   \\ \hline 
 NS5& $\otimes$ & $\otimes$  &$\otimes$  & $\stackrel{z=0}{\bullet}$ & & & & $\otimes$ & $\otimes$ & $\otimes$ \\ \hline 
D3& $\otimes$ & $\otimes$  &$\otimes$  & $\stackrel{z<0}{\otimes}$       &     &     &     &   &   &   \\ \hline
\end{tabular}
    \caption{The zoom-in on the branes sourcing the putative KKLT scale-separated AdS$_4$ vacuum.  }
    \label{tab:D5-NS5-D3_config}
\end{table}

The backreaction of the D3-D5-NS5 branes above gives rise to a near-brane geometry with an $AdS_4$ factor:
\be
ds^2= f_4^2 ds^2_{\mathrm{AdS}_4}+f_1^2 ds^2_{S_1^2}+f_2^2 ds^2_{S_2^2}+4\rho^2 |dw|^2 \,,
\ee
 where $w$ parameterizes a Riemann surface,  $\Sigma_2$, with the topology of a disc \cite{DHoker:2007zhm,DHoker:2007hhe,Aharony:2011yc,Assel:2011xz}. 
The warp factors $f_1$, $f_2$, $f_4$ are controlled by two harmonic functions on $\Sigma_2$, $h_1$ and $h_2$, whose poles determine the location of the D5 and NS5 branes and the number of D3 branes ending on them \cite{DHoker:2007zhm,DHoker:2007hhe,Aharony:2011yc,Assel:2011xz}. Note that these solutions describe a near-brane limit of semi-infinite D3 branes that end on and pull the D5 and NS5 branes into a mohawk-like structure \cite{Bena:2024dre}.

As before we can determine the number of degrees of freedom of the UV theory near these branes by computing the radius of the UV AdS$_4$ in four-dimensional Planck units
\be
\frac{l_{\mathrm{AdS}_4}}{G_\mathrm{N}^{(4)}} = \frac{V_6}{G_\mathrm{N}^{(10)}} \,,
\ee
where $V_6$ is volume of the compact space $\rm S_1^2 \times S_2^2 \times_\mathrm{w} \Sigma_2$, given by \cite{Assel:2012cp,Karch:2022rvr}
\be
V_6 = 2^9 \pi^{2} \int_{\Sigma_2} d^2 x \, h_1 h_2  \left| \partial \bar{\partial}\( h_{1} h_{2}\) \right| \,.
\ee

To make the argument simpler and to relate to the analysis in \cite{Assel:2012cp} we assume the numbers of D5 and NS5 branes to be the same, $N_{\rm D5} = N_{\rm NS5} =N$. Tadpole cancelation in the compact CY manifold requires that the number of D3 branes be of order the product of these numbers $N_{\rm D3} = \alpha N^2$, with $\alpha$ a number of order one that depends on the intricate details of the compactification.

If the distance in the Riemann surface between the poles corresponding to the D5 and NS5 branes is $2 \delta$, the D3 charge and the 5-brane charges are related by \cite{Assel:2012cp} 
\be
N_{\rm D3} = \frac{2 N^2}{\pi} \arctan e^{2 \delta} \,,
\ee
and the volume of the internal space is 
\be
V_6 = G_{\rm N}^{(10)} {4 N^4 \delta e^{-4\delta} \over \pi^2}\,.
\ee
This expression goes to zero for $\delta = 0$ or $\delta \rightarrow \infty$, corresponding to $N_\mathrm{D3} = N^2/2$ and $N_\mathrm{D3} = N^2$.
Hence, it has a maximum somewhere in between (at $\delta = 1/4$) where it takes the value $V_6 = \frac{G_{\rm N}^{(10)}}{e \pi^2} N^4$.
 
This establishes that UV AdS$_{4}$ radius expressed in Planck units scales at most like
\be \label{Free_energy_IIB}
\frac{l^{\rm UV}_{\mathrm{AdS}_4}}{G_\mathrm{N}^{(4)}} \lesssim N^4 \,.
\ee
This places an upper bound on the radius of the putative KKLT scale-separated AdS$_{4}$ in the IR.

It is important to clarify whether the domain walls we construct can give rise to a static solution. Even if the fluxes that give rise to the KKLT solution are supersymmetric, the branes that source them may not be wrapping a Special Lagrangian submanifold, and may break supersymmetry. This happens for example with the choice of fluxes in \cite{Demirtas:2019sip,Demirtas:2021nlu}. Naively, this could result in the destabilization of the domain wall and the invalidation of our arguments. To see that this does not happen it is easiest to consider an example where the  CY three-fold is replaced by a six-torus. The D5 branes of this domain wall would be wrapped on the $123$, $145$, $246$ and $356$ cycles, and the corresponding  NS5 branes would be wrapped on the dual cycles. If the D5 charges $Q_{123}, Q_{145}, Q_{246}, Q_{356}$ have the same sign, the domain wall is supersymmetric. However, if one of these charges are flipped, the supersymmetry is broken.\footnote{
 The D5 branes alone are T-dual to a D4-D4-D4-D0 system, where exactly the same phenomenon happens: if one flips the charge of the D0 branes or of one of the D4 branes, supersymmetry is broken. Such branes are known as almost-BPS  \cite{Goldstein:2008fq, Bena:2009ev}.}

It is easy to see that two such domain walls feel no force between them, and give rise to a static solution, despite the absence of supersymmetry. Supersymmetry is only broken when {\em all} four types of branes are present, and all one-to-one interactions between the branes in one domain wall and the branes in the other are the same as if the domain walls were BPS. This no-force phenomenon is responsible for the existence of the whole plaethora of almost-BPS solutions \cite{Bena:2009ev}. 

\vspace*{-.75cm}
\section{Conclusion}
\vspace*{-.45cm}

In this Letter we have computed the upper bound on the number of degrees of freedom of the  brane systems that source putative AdS vacua with scale separation in M-theory and Type IIB String Theory.

The highest number of degrees of freedom these branes can carry, and hence the weakest bound on the AdS scale separation is achieved if all these branes intersect at only one point within the Calabi-Yau. More generic intersections, in which the branes wrapping the Calabi-Yau manifolds intersect at sparsely distributed points will have less degrees of freedom. 
In the extreme limit, where the number of intersection points is maximized and only one pair of branes is intersecting at each point, our arguments imply that number of degrees of freedom sitting at each intersection is of order one, and hence 
the total number of degrees of freedom scales like the intersection number of the brane cycles, $N^2$, reproducing the result of \cite{Lust:2022lfc}.

On the other hand, if all intersection points are brought together to the same point, we expect additional degrees of freedom. The near-brane AdS solutions we find  capture all these degrees of freedom, and indicate that the total numbers of degrees of freedom can scale like $N^3$ for AdS$_3$ \eqref{eq:Mtheoryc}  and like $N^4$ for AdS$_4$ \eqref{Free_energy_IIB}.
Restricting to the local geometry near the point where the branes intersect has allowed us to ignore the complicated embedding of the branes into the compact Calabi-Yau manifold and obtain general bounds which are independent of these details.

Even if these bounds are parametrically weaker than those of \cite{Lust:2022lfc}, they are still strong enough to rule out KKLT-like AdS vacua with exponential scale separation. Hence, our result indicate that the embedding of our Universe in String Theory cannot take the ``traditional'' form, with an exponential hierarchy between the size of the compactification manifold and the size of the curvature radius of the Universe. Our work highlights the need to start constructing  compactifications with weaker scale separation (such as \cite{Montero:2022prj}) that are still compatible with phenomenological constraints.

\noindent
{\bf Acknowledgements.}
We would like to thank Costas Bachas, Nikolay Bobev, Elias Kiritsis, Manki Kim, Jakob Moritz, Andreas Schachner, Christoph Uhlemann, Cumrun Vafa, Alexander Westphal, and Max Wiesner for interesting discussions.
The work of IB was supported in part by the ERC Grants 787320 ``QBH Structure'' and 772408 ``Stringlandscape'' and by the NSF grant PHY-2309135 to the Kavli Institute for Theoretical Physics (KITP). The work of YL is supported by the University of Padua under the 2023 STARS Grants@Unipd programme (GENSYMSTR – Generalized Symmetries from Strings and Branes) and in part by the Italian MUR Departments of Excellence grant 2023-2027 ``Quantum Frontiers'', and was supported by the German Research Foundation through a German-Israeli Project Cooperation (DIP) grant ``Holography and the Swampland''.


\bibliography{KKLTexnihilo}

\end{document}